\documentclass[aps,pra,floatfix,twocolumn,superscriptaddress]{revtex4-1}

\usepackage{amsmath}
\usepackage{graphicx}
\usepackage{ulem}
\usepackage{bbm}
\usepackage{bm}
\usepackage{color}
\usepackage{mathrsfs}
\usepackage{amsfonts}

\usepackage{hyperref}
\hypersetup{backref,pdfpagemode=FullScreen,colorlinks=true,breaklinks,urlcolor=blue,linkcolor=blue,citecolor=blue}

\begin{document}

\title{Interaction induced topological Bogoliubov excitations in a spin-orbit coupled Bose-Einstein condensate}

\author{Guan-Hua Huang}
\affiliation{Department of Physics, Southern University of Science and Technology, Shenzhen, 518055, China}

\author{Guang-Quan Luo}
\affiliation{Department of Physics, Southern University of Science and Technology, Shenzhen, 518055, China}

\author{Zhigang Wu}
\email{wuzg@sustech.edu.cn}
\affiliation{Shenzhen Institute for Quantum Science and Engineering, Southern University of Science and Technology, Shenzhen, 518055, China}

\author{Zhi-Fang Xu}
\email{xuzf@sustech.edu.cn}
\affiliation{Department of Physics, Southern University of Science and Technology, Shenzhen, 518055, China}
\affiliation{Shenzhen Institute for Quantum Science and Engineering, Southern University of Science and Technology, Shenzhen, 518055, China}

\begin{abstract}
We study topologically non-trivial excitations of a weakly interacting, spin-orbit coupled  Bose-Einstein condensate in a two-dimensional square optical lattice, a system recently realized in experiment [W. Sun et al., Phys. Rev. Lett. {\bf 121}, 150401 (2018)]. We focus on situations where the system is not subjected to a Zeeman field and thus does not exhibit nontrivial single-particle band topology. Of special interest then is  the role of particle interaction as well as its interplay with the symmetry properties of the system in producing topologically non-trivial  excitations. We find that the non-interacting system possesses a rich set of symmetries, including the $\mathcal{PT}$ symmetry, the modified dihedral point group symmetry $\tilde D_4$ and the nonsymmorphic symmetry. These combined symmetries ensure the existence of pairs of  degenerate Dirac points at the edge of Brillouin zone for the single-particle energy bands. In the presence of particle interaction and with sufficient spin-orbit coupling, the atoms condense in a ground state with net magnetization which spontaneously breaks the $\mathcal{PT}$ and $\tilde D_4$ symmetry. We demonstrate that this symmetry breaking leads to a gap opening at the Dirac point for the Bogoliubov spectrum and consequentially topologically non-trivial excitations. We confirm the non-trivial topology by calculating the Chern numbers of the lowest excitation bands and show that gapless edge states form at the interface of systems characterized by different values of the Chern number. 
\end{abstract}

\maketitle
\section{Introduction}
\label{Introduction}
Simulating topological phases of matter using cold atomic gases has seen tremendous progresses in recent years~~\cite{Hofstadter_Hamiltonian,Harper_Hamiltonian,Haldane_2014,TopologicalBandColdAtom_RMP,atala2013direct,Lohse2016,Nakajima2016,Leder2016,meier2016observation,xie2019topological,leseleuc2019}. For example, seminal models  of topological insulators such as the Haldane model~\cite{Haldane_2014} and the SSH model~\cite{atala2013direct,Lohse2016,Nakajima2016,Leder2016,meier2016observation,xie2019topological,leseleuc2019} have been realized in experiments and their topological properties were investigated. Such realizations are made possible because of the many experimental tools available to generate desired lattice potentials and to engineer complex valued site-to-site tunneling for the atoms, two crucial ingredients for creating topologically non-trivial energy bands. 

Another component that is of great significance to topological phases of matter is spin-orbit (SO) coupling~\cite{BCS-BEC.Crossover,Unconventional.soc.CongjunWu,Realization.2D.XJLiu,Weyl_FermiSuperfluids,Chen_Shuai_2016realization,Observation_Gap_Fermi_ZhangJ,Yi_Wei_PhysRevA.94.043619,SPT_Observation,Sun_Wei_Experiment,Xiong-JunLiu-3D}. In fact, for condensed matter systems the intrinsic SO coupling  has long been recognized to play an essential role in topological insulators~\cite{Topological_RevModPhys.82.3045}, topological superconductors~\cite{RevModPhys_TopSuperconductors} and topological semimetals~\cite{Weyl_Semimetal,Weyl_Semimetal_TaAs}. 
Thus, recent realizations of two-dimensional (2D) synthetic SO coupling using spacially dependent Raman potential~\cite{Fermi_gases_ZhangJ,Observation_Gap_Fermi_ZhangJ,Chen_Shuai_2016realization,Sun_Wei_Experiment} have further added to the toolbox of cold atom experimentalists in their ability to explore topological phases. Indeed, non-trivial band topology has already been demonstrated for both Bose and Fermi gases with 2D   
synthetic SO coupling~\cite{Chen_Shuai_2016realization,Sun_Wei_Experiment,Observation_Gap_Fermi_ZhangJ}. In these and many other earlier studies of topological bands, the focus is on non-interacting systems where the band gap at the Dirac point is opened either by introducing complex valued tunnelings or a Zeeman field;  particle interactions in these systems do not play a relevant role in  non-trivial band topology.  

In some cases, however, particle interactions play an important role in creating non-trivial topology for both fermion~\cite{Topological.Mott.Insulators,sunkai.2009,sunkai_topological_2011,PhysRevA.86.053618.2012,Liu2014,PhysRevA.93.043611.2016,Odd-parity.topological.superfluidity} and boson~\cite{inversion-symmetric,Furukawa_2015,ChiralBosonic.Nonlinear,Dirac.Bosons,Varma.Superfluid,peano2016.topological,Bosonic.top.excitations,PhysRevB.87.174427,Zhou_2020,Flynn_2020_number_conservation,PhysRevA.102.043323,wan2020squeezinginduced}. Chiral $p$-wave superfluids are one example where pairing interaction of $p$-wave symmetry is essential. However, it is very challenging to realize chiral $p$-wave superfluids in cold atomic systems and this is yet to be done  despite the existence of many theoretical proposals~\cite{PhysRevLett.101.160401,NISHIDA2009897,Liu2014,PhysRevLett.117.245302,PhysRevA.94.063631,PhysRevA.96.033605}.  SO coupled Bose gases in a 2D optical lattice is another type of systems for which non-trivial topological bands of collective excitations may emerge in the presence of particle interactions~\cite{Yi_Wei_PhysRevA.94.043619}. The single-particle energy bands of 2D SO coupled gases are known to feature Dirac points which are protected by certain underlying symmetries of the system. Introducing external fields may explicitly break such symmetries and lead to topologically non-trivial excitation bands, but this is not the only way. The main purpose of this paper is to provide an in-depth analysis of how particle interactions in these systems can also break such symmetries  in a spontaneous way and in doing so create topologically non-trivial excitations.  

Our system of interest is a weakly interacting, SO coupled  Bose gas in a 2D square optical lattice potential, which has been realized in a recent experiment~\cite{Sun_Wei_Experiment}.  In Sec.~\ref{model}, we determine the  non-interacting band structure for this system and analyze the symmetry properties responsible for its most essential properties. Interestingly, the system does not have the usual $D_4$ point group symmetry as one might assume. Rather, as a result of the SO coupling, it has an enlarged point group symmetry which is precisely a double group of $D_4$. We show that this group symmetry, $\tilde D_4$, plays a crucial role in protecting the stability of the Dirac points in the non-interacting band structure. In Sec.~\ref{Ground states}, we take into account the particle interactions and determine the condensate wave function. We show that the ground state breaks the $\tilde D_4$ symmetry spontaneously by analyzing the spin configuration of the condensate wave function.  The consequence of this symmetry breaking is explored in Sec.~\ref{excitation} where we calculate the collective excitations and demonstrate the non-trivial excitation band topology. All the results are summarized in finial section~\ref{Conclusion}.

\section{2D spin-orbit coupling model}
\label{model}
\begin{figure}[htbp]
	\centering
	\includegraphics[width=8.4cm]{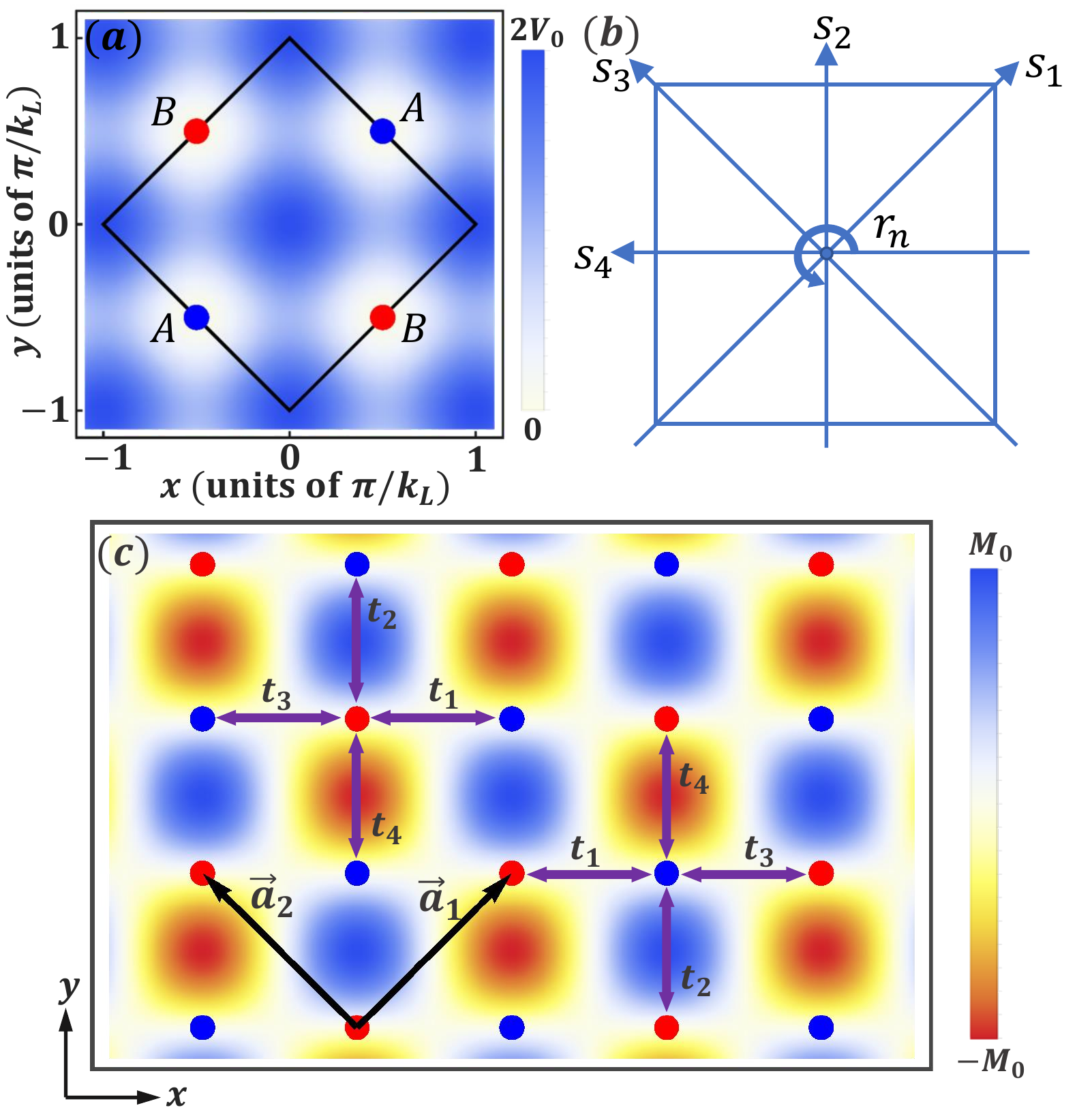}
	\caption{(a) The  Wigner-Seitz cell of the lattice system. The false color in the background indicates the optical lattice potential. (b) The dihedral point group symmetry $D_4$ for the single particle Hamiltonian in the absence of the Raman potential. (c) The $A$ and $B$ sub-lattices are distinguished by the couplings to their nearest neighboring sites due to the Raman potential. The false color in the background indicates the  part of the Raman potential proportional to $\sigma_x$.} 
	\label{fig:image0}
\end{figure}

We consider a 2D Bose gas in an optical lattice in which a synthetic SO coupling has been realized in a recent experiment by means of a clever Raman scheme~\cite{Sun_Wei_Experiment}. The single-particle Hamiltonian of this system is given by
\begin{equation}
	\label{eq:H0}
	h_0=\frac{\bm p^2}{2m}+V_{\text{latt}}(x,y)+V_R(x,y),
\end{equation}
where $V_{\text{latt}}(x,y)=V_0(\cos^2k_{L}x+\cos^2k_{L}y)$ is the 2D optical potential formed by two lasers with wavelength $\lambda = 2\pi/k_L$ and $ V_{R}(x,y)=M_0\sin (k_{L}x)\cos (k_{L}y)\sigma_{x}+M_0\sin (k_Ly)\cos (k_{L}x)\sigma_{y}$ is the Raman potential that couples two hyperfine states of the atom.  Here, $V_0$ and $M_0$, tunable in experiments, are depths of the lattice  and Raman potential respectively, and $\sigma_{x(y)}$ are Pauli matrices. In the absence of SO coupling, the optical potential forms a 2D square lattice shown in Fig.~\ref{fig:image0}(a), which has the usual $D_4$ symmetry. However, the SO coupling divides the lattice into two sublattices, because the  nearest neighbor couplings within the $A$ and $B$ sublattices are in fact  antisymmetric. This can be inferred from Fig.~\ref{fig:image0} (a) and (c). 

The SO coupling considered here is a variant of an earlier one which was also realized in experiments~\cite{Chen_Shuai_2016realization,Yi_Wei_PhysRevA.94.043619}. However, the symmetry properties of these two models are drastically different, which have important implications with regard to the band topology. In particular, as we shall explore in this section, the richer symmetry exhibited in the model considered here leads to a  non-interacting band structure with robust Dirac points.
\subsection{Band structure and Dirac points}
\begin{figure}[htbp]
	\centering
	\includegraphics[width=8.8cm]{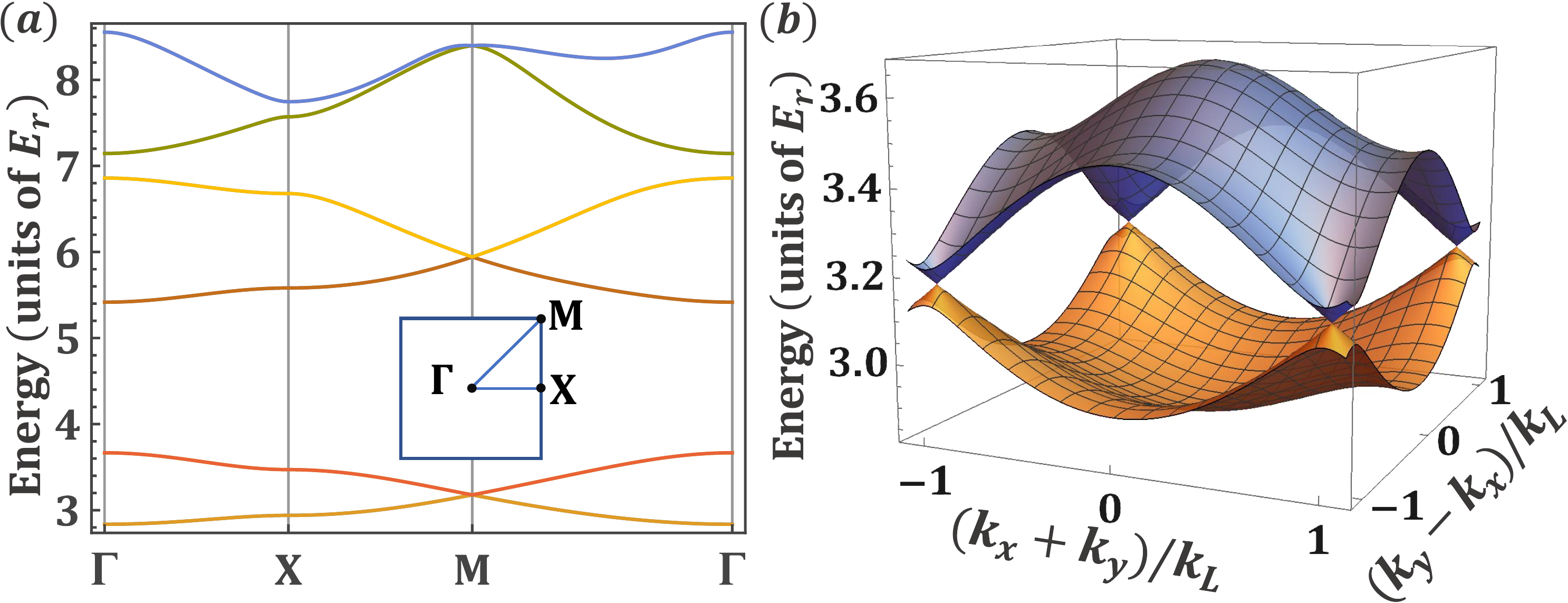}
	\caption{(a) The lowest several non-interacting energy bands along the high symmetry points in the first Brillouin zone.  (b) The lowest two bands in the first Brillouin zone.   Each band is doubly degenerate and the Dirac points arise at the $M$ points. Here and for the rest of the paper, $V_0=5.2E_r$ and $M_0=3.0E_r$, where $E_r=\hbar^2k_L^2/2m$ is the recoil energy }
	\label{fig:img1}
\end{figure}

The band structure of the single-particle Hamiltonian in Eq.~(\ref{eq:H0}) can be computed straightforwardly from a plane wave expansion. Although one may construct a four-band model based on the tight-binding approximation in the deep lattice limit $V_0\gg M_0$, earlier work on related models~\cite{Yi_Wei_PhysRevA.94.043619} has shown that this may lead to qualitatively incorrect band structures when the strength of the SO coupling becomes significant. The plane wave expansion in comparison yields a very accurate band structure. For our calculation, the primitive vectors are chosen to be $ \bm a_1=(1,1)\frac{\pi}{k_{L} }$ and $\bm a_2=(-1,1)\frac{\pi}{k_{L}}$, from which the reciprocal lattice vectors $\bm b_j$ can be determined via $\bm a_i\cdot\bm b_j=2\pi\delta_{ij}$. As an example, we show in Fig.~\ref{fig:img1} the lowest few  bands calculated for $V_0 = 5.2 E_r$ and $M_0 = 3 E_r$, where $E_r = \hbar^2 k^2_L/2m$ is the recoil energy. The values of these parameters are comparable to those in the recent experiment~\cite{Sun_Wei_Experiment} and will be used throughout this work. 
 From our numerical results for various system parameters, we can establish the following general properties for the band structure:
 (i) Each band is doubly degenerate; (ii) A pair of degenerate Dirac points appear at the $M$ point of the Brillouin zone for the lowest four bands; and (iii) The location of these Dirac points does not change when the depths of the lattice or Raman potential are varied. As we shall argue below, these properties are consequences of various symmetries possessed by the Hamiltonian $h_0$. 
\subsection{Symmetry analysis}
\label{sa}
The global stability of the Dirac points at the $M$ point is protected by the $\mathcal{P}\mathcal{T}$ symmetry, the modified dihedral symmetry $\tilde D_4$ together with the so-called nonsymmorphic symmetry. Although $h_0$ does not separately have the inverse or time-reversal symmetry, it can be checked that it has the combined $ \mathcal{P}\mathcal{T}$ symmetry. $\mathcal{P}\mathcal{T}$ is an anti-unitary operator and satisfies ($\mathcal{\mathcal{P}\mathcal{T}})^2=-1$. In addition, it relates two orthogonal Bloch wave functions with the same quasi-momentum $\bm k$. As a result, we have the double degeneracy of each band, analogous to the Kramers degeneracy in spin-$1/2$ electron systems. 
\begin{table}[htbp]
	\centering
	\[
	\begin{array}{c|ccccccc}
	{\tilde D_4}  & \{e\} & \{\tilde r_4\} & 2\{\tilde r_1\} & 2\{\tilde r_3\} & 2\{\tilde r_2\} & 4\{\tilde s_1\} & 4\{\tilde s_2\} \\
	\hline
	A_1 & 1 & 1 & 1 & 1 & 1 &1 & 1 \\
	A_2 & 1 & 1 & 1 & 1 & 1 & -1 & -1\\
	B_1 & 1 & 1 & -1&-1 &1 & -1 & 1 \\
	B_2 & 1 & 1 & -1&-1 & 1 & 1 & -1 \\
	E_1 & 2 & 2 & 0 & 0 &-2 & 0 & 0 \\
	E_2 & 2 & -2& \sqrt{2}&-\sqrt{2}& 0 & 0 & 0 \\
	E_3 & 2 & -2& -\sqrt{2}&\sqrt{2}& 0 & 0 & 0 
	\end{array}
	\]
	\caption{The character table of $\tilde D_4$.}
	\label{D4character}
\end{table}

In the absence of the SO coupling, the system has the usual point group $D_4$ symmetry, which consists of the four-fold rotation operations $r_n$ and the two-fold reflection operations $s_n$  ($n = 1,\cdots,4$) illustrated in Fig.~\ref{fig:image0} (b). Here $r_n$ denotes the counterclockwise rotation of $n\pi/2$ around the $z$-axis at the origin and $s_n$ denotes the reflection across a line that makes an angle of $n\pi/4$ with the $x$-axis.  With SO coupling, however, the Hamiltonian $h_0$ no longer commutes with these  operations. Instead, it commutes with what we will refer to as the modified dihedral symmetry operations
\begin{align}
\tilde{r}_n\equiv  e^{-i\frac{n\pi}{4}\sigma_z}r_n\qquad \tilde{s}_n\equiv  e^{-i\frac{\pi}{2}\vec s_n \cdot \vec \sigma}s_n
\end{align}
for $n = 1,\cdots,4$ and 
\begin{align}
\tilde{r}_n\equiv  e^{-i\frac{n\pi}{4}\sigma_z}r_{n-4}\qquad \tilde{s}_n\equiv  e^{i\frac{\pi}{2}\vec s_{n-4} \cdot \vec \sigma}s_{n-4}
\end{align}         
for $n = 5,\cdots,8$. Here $\vec s_{n}$ is the unit vector along the reflection axis of the $s_n$ operation. We find that to maintain the invariance of the Hamiltonian during the $D_4$ symmetry operation, the spin degrees of freedom need to be rotated at the same time. These $16$ operations form a symmetry group of $h_0$ which is  a double group of $D_4$ point group \cite{Group_Cornwell,dresselhaus2007group} and whose character table is shown in Table.~\ref{D4character}. This symmetry group, denoted by $\tilde D_4$, is essential in protecting the Dirac points. To demonstrate this we analyze the Bloch states at the highly symmetric point in the Brillouin zone, i.e., the $M$ point. 
The lowest four degenerate eigenstates at the $M$ point $\{\phi_{M1},\phi_{M2},\phi_{M3},\phi_{M4}\}$ form the basis of a four-dimensional representation $E_M$ of the $\tilde D_4$ group. We calculate the characters of $E_M$ numerically and obtain Tabel.~\ref{D4tabel}.
\begin{table}[htbp]
\centering
\[\begin{array}{c|ccccccc}
	{\tilde D_4}  & \{e\} & \{\tilde r_4\} & 2\{\tilde r_1\} & 2\{\tilde r_3\} & 2\{\tilde r_2\} & 4\{\tilde s_1\} & 4\{\tilde s_2\} \\
	\hline
	E_M & 4 & -4 & 0 & 0 & 0 & 0 & 0
	\end{array}\parbox[c][2em][b]{1em}{.}
\]
\caption{Numerically calculated character of $E_M$.}
\label{D4tabel}
\end{table}
From Tabel.~\ref{D4tabel}, we find that $E_M=E_2\oplus E_3$. That is to say the four eigenstates at the $M$ point can be further divided into two subsets $\{\phi_{M1},\phi_{M2}\}$ and $\{\phi_{M3},\phi_{M4}\}$, each forming the basis of a 2D irreducible representation of the $\tilde D_4$ group.  It turns out that these two subsets can in fact be transformed from one to another under the so-called nonsymmorphic symmetry operations~\cite{Zhou_Qi_PhysRevA.95.053615}. These symmetry operations are described by $T_{\hat x(\hat y)}(\frac{\pi}{k_L})\exp(-i\frac{\pi}{2}\sigma_{z})$, where $T_{\hat x(\hat y)}(l)$ is a translation along the $x(y)$-direction for a distance $l$.  It is confirmed that the nonsymmorphic symmetry operations commutes with $h_0$, which implies that $\tilde D_4$ group symmetry together with nonsymmorphic symmetry lead to the the four-fold degeneracy at the $M$ point. In view of the full two-fold degeneracy due to the $\mathcal{PT}$ symmetry, the four-fold degeneracy indicates that  $M$ point is in fact a band-touching point. Furthermore, the linear dispersion in the vicinity of the $M$ point, namely the fact that $M$ point is a Dirac point, can be derived from a perturbative method combined with the $\tilde D_4$ group symmetry analysis (see  Appendix~\ref{appendixA}). Consequently, breaking the $\tilde D_4$ symmetry will remove the four-fold degeneracy at the $M$ point and open a gap at the Dirac point.

\section{Ground state: Spontaneous  breaking of point group symmetry $\tilde D_4$}
\label{Ground states}
So far we have only analyzed the single-particle spectrum of the SO  coupled system, which itself does not exhibit non-trivial band topology. To set the stage for studying interaction induced topological excitations,  we turn to the ground state of an interacting Bose gas described by the Hamiltonian 
\begin{align}
\label{H}
\hat H =& \sum_{\sigma\sigma'}\int d\bm r\, \hat\psi^{\dagger}_{\sigma}(\bm r)h_{0,\sigma \sigma'}\hat\psi_{\sigma'}(\bm r)  \nonumber \\
&+\frac{1}{2}\sum_{\sigma\sigma'}g_{\sigma\sigma'}\int d\bm r\, \hat\psi^{\dagger}_{\sigma}(\bm r)\hat\psi^{\dagger}_{\sigma'}(\bm r)\hat\psi_{\sigma'}(\bm r)\hat\psi_{\sigma}(\bm r)
\end{align}
where $\hat\psi^{\dagger}_{\sigma}(\bm r)$ is  the field operator,  $\sigma=\uparrow,\downarrow$ denotes the spin index and $g_{\sigma\sigma'}$ is the interaction strength. For the following calculations, the values of the interaction strengths are chosen to be $g_{\uparrow\uparrow}=g_{\downarrow\downarrow}=0.3497E_r/n_{0}$ and $g_{\uparrow\downarrow}=g_{\downarrow\uparrow}=0.3489E_r/n_{0}$ where $n_0$ is the average number density, which are the values found in a recent experiment~\cite{Sun_Wei_Experiment}.

The bosonic atoms will condense at zero temperature and the condensate wave function $\phi_{0\sigma}(\bm r)\equiv \langle \hat \psi_\sigma (\bm r)\rangle$ can be calculated by minimizing the Gross-Pitaevskii energy functional. More explicitly, we express the trial wave function in terms of the superposition of single-particle wave functions at the $\Gamma$ point and compute the coefficients using a global optimization method known as the simulated annealing~\cite{Simulated.Annealing}. Consistent with previous studies~\cite{Sun_Wei_Experiment}, we find that the system is in the stripe phase for sufficiently small SO coupling strength, where the atoms are in coherent superpositions of spin-up and spin-down state. Beyond  a critical SO coupling strength, the system enters the magnetic phase (or plane wave phase), where the Bose gas exhibits a net magnetization. We focus on the latter phase because it provides an opportunity to study  topological excitations. 

\begin{figure}[htbp]
	\centering
	\includegraphics[width=8.8cm]{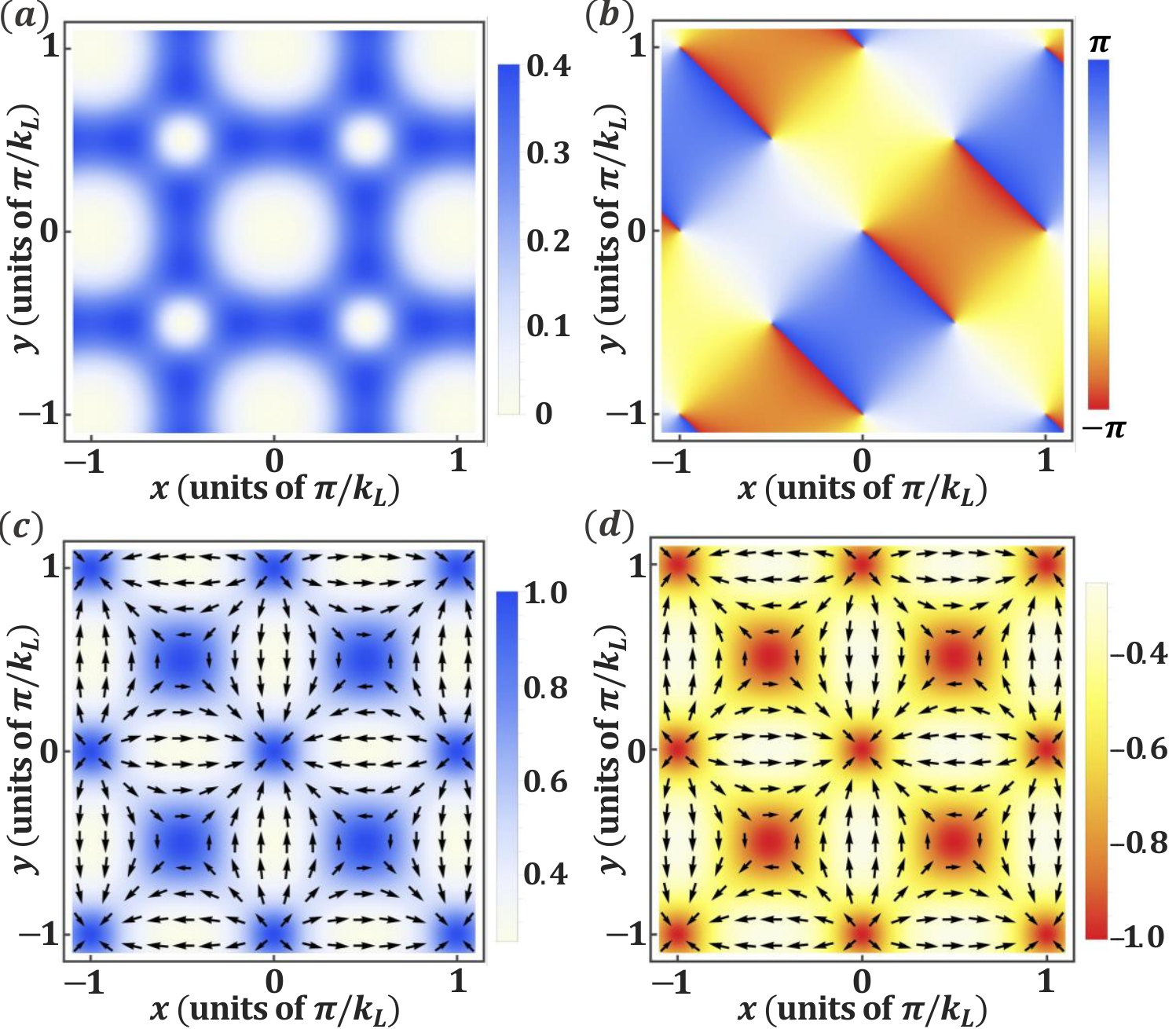}
	\caption{(a) The probability density distribution of the spin down atoms in the ground state. (b) The phase of the wave function for the spin down atoms.   (c) The spin orientation for the condensate wave function $\phi_0$.  (d) The spin orientation for the condensate wave function $\tilde\phi_0 = s_1[\phi_0]$. The color bars in (c) and (d) indicate the strength of the $z$-component of the spin-orientation vector.}
	\label{fig:img2}
\end{figure}

In the magnetic phase, the Bose gas chooses either one of two degenerate ground states, which have magnetization of opposite direction and are $\mathcal{P}\mathcal{T}$ symmetric with respect to each other. Once the condensate wave function is chosen, various physical quantities such as the particle densities and spin orientations can be readily calculated.  To be specific, we consider the ground state with the magnetization along the $z$-direction, where the majority of atoms are in the spin up state. The first notable aspect about the condensate wave function is the significant mixing of the $p$-orbital state of the lattice potential.  In Fig.~\ref{fig:img2} (a) and (b) we show the probability density and the wave function phase for the spin down component within a primitive cell. The $2\pi$ phase winding around the lattice sites clearly shows the $p$-wave nature of the spin down atoms.  Because the nature of the condensate wave function for a weakly interacting gas largely reflects that of the single-particle ground state, this is an indication that the tight-binding model involving only lowest $s$-orbitals is not sufficient to determine the single-particle band structure.

 The second and more important aspect of the condensate wave function is  that it spontaneously breaks the $\tilde D_4$ symmetry. More specifically, it breaks the reflection symmetry $\tilde s_n$ while still preserving the rotational symmetry $\tilde r_n$. To see this we plot in Fig.~\ref{fig:img2} (c) the local spin orientation vector defined as
\begin{align}
  \bm m(\bm r)=\frac{1}{\phi_0^{\dagger}(\bm r)\phi_0(\bm r)}\sum_i \phi_0^{\dagger}(\bm r)\sigma_{i}\phi_0(\bm r)\bm e_i,
  \end{align}  
  where $\sigma_{i=x,y,z}$  is the Pauli matrix and  $\phi_0(\bm r)\equiv\left[\phi_{0\uparrow}(\bm r)\,\,\phi_{0\downarrow}(\bm r)\right]^{T}$. We can also introduce a new spin orientation  $\tilde {\bm m} (\bm r)$ defined in terms of the transformed condensate wave function
  \begin{align}
   \tilde {\bm m}(\bm r)=\frac{1}{\tilde \phi_0^{\dagger}(\bm r)\tilde\phi_0(\bm r)}\sum_i \tilde \phi_0^{\dagger}(\bm r)\sigma_{i}\tilde\phi_0(\bm r)\bm e_i,
  \end{align}  
  where $\tilde \phi_0 \equiv \tilde D_4 [\phi_0]$ is the wave function transformed by one of the symmetry operations in $\tilde D_4$.
It is straightforward to show that 
\begin{align}
\tilde {\bm m} (\bm r) = D_4 [\bm m (\bm r)].
\end{align}
This is to say that the new spin orientation can be obtained simply by performing the corresponding symmetry operation in $D_4$ on $\bm m (\bm r)$. For example, we show in Fig.~\ref{fig:img2} (d) the spin orientation for $\tilde  \phi_0 = s_1[\phi_0]$. By inspection, we can see that the spin orientation retains the four-fold rotational symmetries but violates the two-fold reflection symmetries, reflecting the fact that  the condensate wave function breaks the $\tilde D_4$ group symmetry. The consequence of such a spontaneous symmetry-breaking in relation of the topological properties of the system will be explored in the next section. 

\section{Interaction induced topological EXCITATIONS}
\label{excitation}
After solving the ground state of the SO coupled Bose gas, we are in a position to examine the collective excitations. For this purpose, we employ the well-known Bogoliubov theory and write the field operator as
\begin{equation}
	\label{meanfield}
	\hat\psi_{\sigma}(\bm r)=\phi_{0\sigma}(\bm r)+\hat\delta_\sigma(\bm r)
\end{equation}
where $\phi_{0\sigma}(\bm r)$ is  the  condensate wave function  and $\hat\delta_{\sigma}(\bm r)$ is fluctuation operator. With Eq. (\ref{meanfield}), it is straightforward to obtain the Bogoliubov-de Gennes (BdG) Hamiltonian
\begin{equation}
		\mathcal{H}_{BdG}=\frac{1}{2}\int\bm \delta^{\dagger}\begin{pmatrix} \mathcal{M}+h_0 -\mu& \mathcal{N} \\ \mathcal{N}^{*} & \mathcal{M}^{*}+h_0^{*}-\mu\end{pmatrix}\bm \delta d\bm r \label{eq:bdgre}
\end{equation}
where  $\bm\delta\equiv\left(\hat\delta_{\uparrow}, \hat\delta_{\downarrow}, \hat\delta_{\uparrow}^{\dagger},\hat\delta_{\downarrow}^{\dagger}\right)^{T}$ and $\mu$ is the chemical potential, 
\begin{equation}
	\mathcal{M}=\begin{pmatrix} 2g_{\uparrow\uparrow}|\phi_{0\uparrow}|^{2}+g_{\uparrow\downarrow}|\phi_{0\downarrow}|^{2} & g_{\uparrow\downarrow}\phi_{0\downarrow}^{*}\phi_{0\uparrow} \\
	g_{\uparrow\downarrow}\phi_{0\uparrow}^{*}\phi_{0\downarrow} & 2g_{\downarrow\downarrow}|\phi_{0\downarrow}|^{2}+g_{\uparrow\downarrow}|\phi_{0\uparrow}|^{2}
	\end{pmatrix} \label{eq:BdgM}
\end{equation}
and
\begin{equation}
	\mathcal{N}=\begin{pmatrix} g_{\uparrow\uparrow}\phi_{0\uparrow}^{2} &g_{\uparrow\downarrow}\phi_{0\downarrow}\phi_{0\uparrow} \\ g_{\uparrow\downarrow}\phi_{0\uparrow}\phi_{0\downarrow} & g_{\downarrow\downarrow}\phi_{0\downarrow}^2 \end{pmatrix}. \label{eq:BdgN}
\end{equation}
Here the value of chemical potential is determined by solving the ground state of the condensate and assume the value $\mu = 3.37 E_r$ for our chosen system parameters. In the following, we diagonalize the BdG Hamiltonian in Eq.~(\ref{eq:bdgre}) under various scenarios so as to explore different aspects of interaction induced topological excitations. 

\subsection{ Excitation spectrum and Chern number}
\begin{figure}[htbp]
	\centering
	\includegraphics[width=8.8cm]{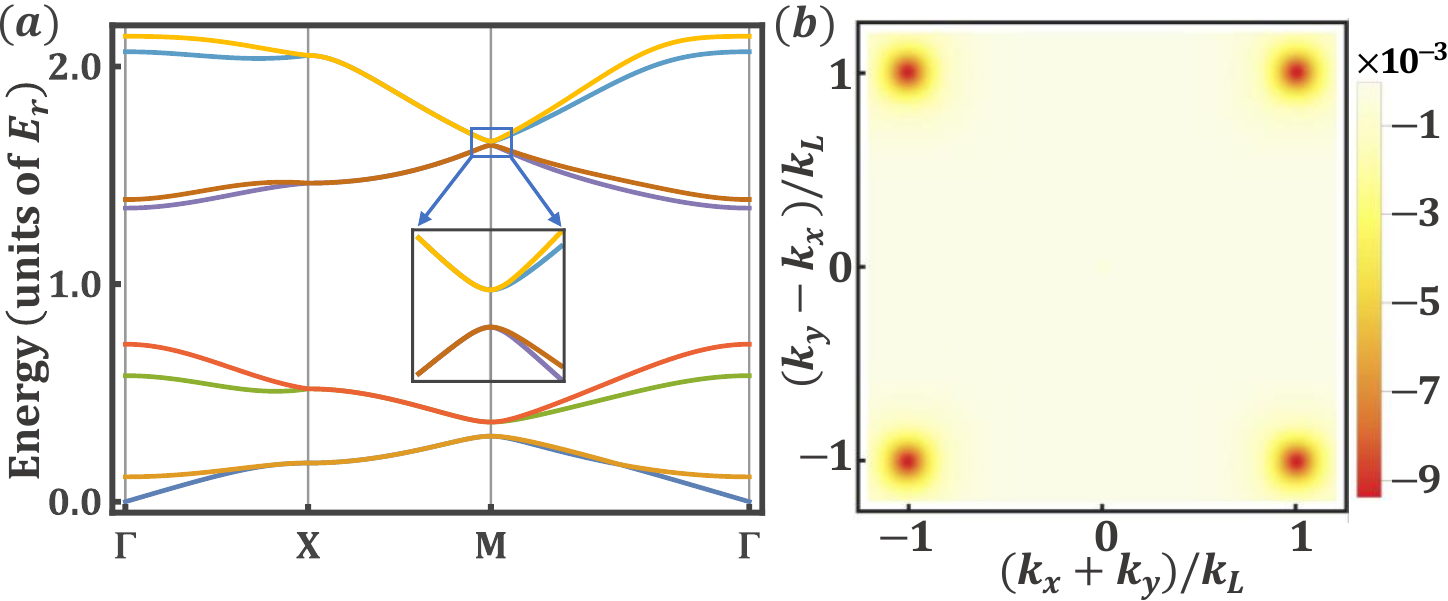}
	\caption{(a) The Bogoliubov excitation spectra of the lowest several bands.  The complete double degeneration of each band is removed and gap openings occur at the Dirac points of the single-particle bands. (b) Berry curvature of the lowest  band. The peaks at the $M$ points all contribute to the non-zero Chern number of the band. }
	\label{fig:img3}
\end{figure}

In Fig~\ref{fig:img3}(a) we show the Bogoliubov spectrum calculated using the condensate wave function obtained in the previous section. Since the ground state spontaneously breaks the $\mathcal{PT}$ and $\tilde D_4$ symmetry, the Bogoliubov Hamiltonian $\mathcal{H}_{\text{BdG}}$  does not have these symmetries. This explains the marked differences between the Bogoliubov spectrum and the single-particle spectrum. First, the breaking of $\mathcal{PT}$ ensures that the full double degeneracy of each band is removed, as can be seen in Fig~\ref{fig:img3}(a). Furthermore, the additional $\tilde D_4$ symmetry breaking leads to the opening of a gap at the Dirac point. We emphasize that $\mathcal{PT}$ symmetry breaking alone does not necessarily mean that a gap will open at the Dirac point. It is conceivable that the pair of degenerate Dirac points may seperate and move away from the $M$ point if $\mathcal{PT}$ symmetry is broken. This is indeed the case for a related model where the non-interacting system does not possess $\mathcal{PT}$ symmetry but still has a pair of non-degenerate Dirac points~\cite{Yi_Wei_PhysRevA.94.043619}. Finally, the Bogoliubov spectrum at the  Brillouin zone boundary ($M$ point) are still two-fold degenerate, because the nonsymmorphic symmetry remains intact for the Bogoliubov Hamiltonian. Thus, the lowest two subbands shown in Fig.~\ref{fig:img3} (a) form one single band and in total we have shown four bands of Bogoliubov excitations in Fig.~\ref{fig:img3} (a).

When a gap opens at the Dirac point as a result of the broken $\tilde D_4$ symmetry, non-trivial band topology for the excitations with a finite Chern number is expected to arise. Since the symmetry breaking is brought by the presence of particle interactions, this type of excitations fall into the broad category of the so-called interaction induced topological excitations~\cite{Zhou_2020}. To demonstrate the non-trivial band topology we calculate the Berry curvature~\cite{PhysRevB.87.174427}
\begin{align}
\Omega(\bm k) =\mathrm{i}\sum_{m,ij}\epsilon_{ij}\partial_{k_i}\langle u_m(\bm k)|\sigma_{z}\partial_{k_j}|u_m(\bm k)\rangle 
\end{align}
and the Chern number
\begin{align}
	C =\frac{1}{2\pi}\int_{BZ}d\bm k\; \Omega(\bm k)
\end{align}
for the lowest several bands of the Bogoliubov excitations. Here $|u_m(\bm k)\rangle$, in the form of spinor, denotes the wave function of the Bogoliubov excitation with a specific magnetization. 
By discretizing the Brillouin zone~\cite{Fukui_2005_ChernNumbers}, we calculate the Berry curvature of the lowest bands in Fig. \ref{fig:img3}(b). Clear peaks emerge at the $M$ points of the Brillouin zone, which together give rise to  a finite Chern number of $-1$. In fact, the second band has opposite Berry curvature to the previous one and so the Chern number alternates between $-1$ and $1$ going from one band to another. 
\subsection{Edge states}
\begin{figure}[htbp]
	\centering
	\includegraphics[width=8.6cm]{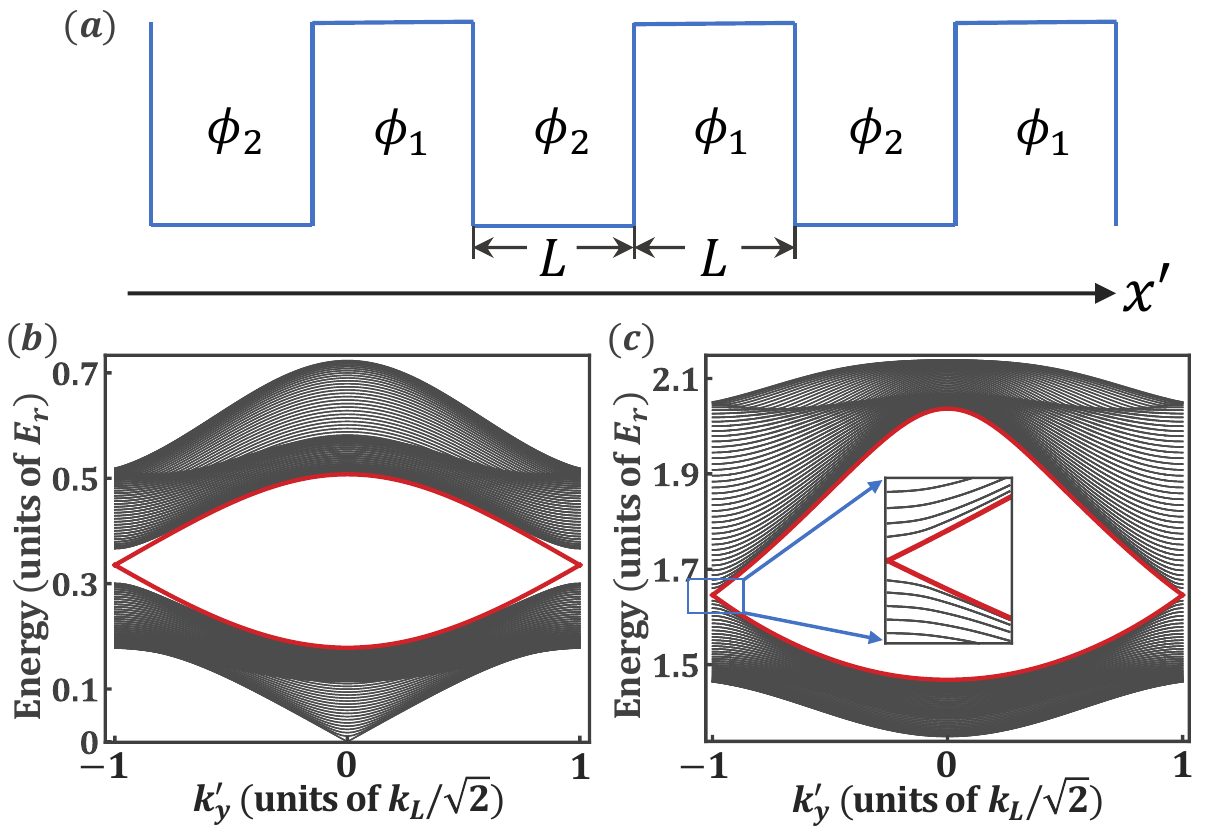}
	\caption{(a) Illustration of the periodic domain walls. (b) The excitation spectrum of the lowest bands for  system with the periodic domain wall. There are clear edge modes which close the gaps at the boundary of the Brillouin zone. (c) The edge modes corresponding to the higher bands.}
	\label{fig:img4}
\end{figure}

To further ascertain the non-trivial band topology of the Bogoliubov excitations, we look for topologically protected edge states formed at the interface separating systems with different Chern numbers~\cite{Topological_RevModPhys.82.3045}. Rather than considering a system-vacuum interface, we construct a set of periodic domain walls along the $\bm a_1$ direction such that the system alternates between two $\mathcal{PT}$ symmetric ground states, as illustrated in Fig.~\ref{fig:img4} (a). More specifically, we consider $N$ copies of the previous system assembled along the $\bm a_1$ direction and calculate the Bogoliubov excitations above the following ground state
\begin{equation}
\phi(\bm r')=
\begin{cases}
\phi_1(\bm r')& (2n-2)L\leq x'<(2n-1)L \\
\phi_2(\bm r') & (2n-1)L\leq x'<2nL
\end{cases}
\label{eq:fx}
\end{equation}
for $n = 1,2,\cdots, N/2$, where $\phi_1(\bm r')$ and $\phi_2(\bm r')$ are two $\mathcal{PT}$ symmetric ground states. Here for convenience we choose the coordinates along the $\bm a_1$ and $\bm a_2$ directions. 
Our plane wave expansion is well-suited to treat such a periodic system. We find that the corresponding Bogoliubov excitation bands above the $\mathcal{PT}$ symmetric ground states have opposite Chern numbers, which explains the formation of gapless edge states shown in Fig.~\ref{fig:img4}
(b) and (c). We have further calculated the real space wave function of these edge states and confirmed that they are well localized at the interfaces.

\subsection{Higher band condensation}
\begin{figure}[htbp]
	\centering
	\includegraphics[width=8.8cm]{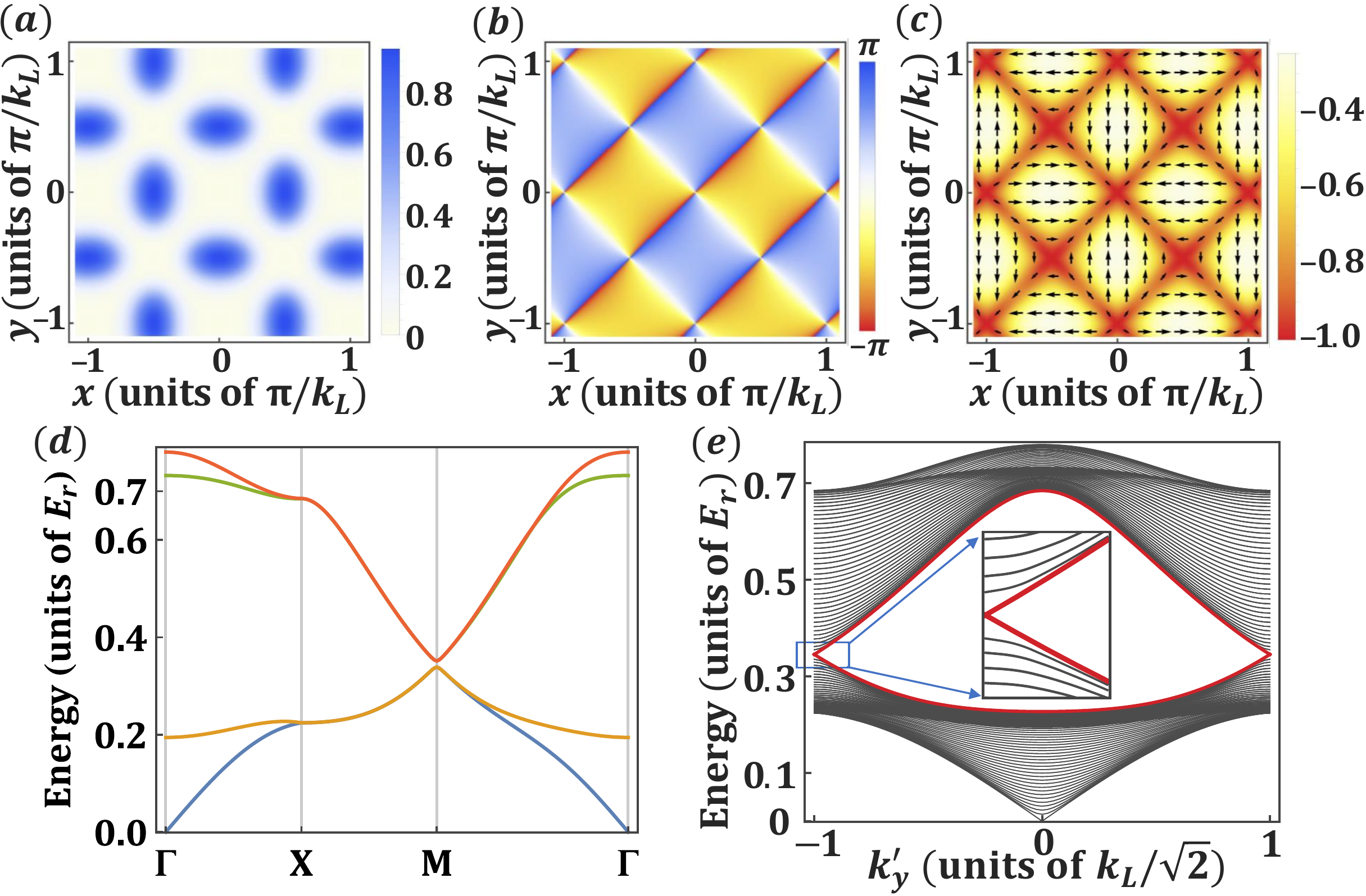}
	\caption{(a) The probability density distribution of the spin up atoms in the new condensate wave function. (b) The phase of the wave function for the spin up atoms.   (c) The spin orientation for the condensate wave function.  (d) The lowest four bands of Bogoliubov excitations. (e) Edge states in the Bogoliubov excitation spectrum of the periodic domain walls.}
	\label{fig:img5}
\end{figure}
The previous analysis on the non-trivial band topology of the Bogoliubov excitations is quite general in the sense that it does not depend on the specific ground state so long as it breaks the $\tilde D_4$ symmetry and $\mathcal{PT}$ sysmetry. To demonstrate this we determine the Bogoliubov excitations assuming that the atoms condense at the local minimum of higher bands. From Fig.~\ref{fig:img1}(a), we see that it is possible to load the atoms onto the $\Gamma$ point of the third  band by tuning the optical lattice. We perform similar calculations on the condensate wave function, the collective excitations, the Chern number and the edge states. The results are summarized in Fig.~\ref{fig:img5}. We note that there are some interesting contrasts between the ground state of the higher band condensate and that of the lower band one considered earlier. Fig.~\ref{fig:img5} (a) and (b) clearly show the $d$-wave nature of the spin up atoms, indicating a significant mixing of the $d$-wave orbitals for the condensate atoms.  In addition, the spin orientation pattern shown in Fig.~\ref{fig:img5} (c) is also drastically different from that in Fig.~\ref{fig:img3}.  Despite these differences in ground states, we see from Fig.~\ref{fig:img5} (d) and (e) that the qualitative features and, in particular, the topological nature of the Bogoliubov excitations are largely the same as those of the lower band condensation.  This confirms our expectations that the Bogoliubov excitations above the higher band condensation are also topologically nontrivial.

\section{Conclusions}
\label{Conclusion}
 In this paper, we consider a 2D SO coupled Bose gas in a lattice potential and address the question of whether particle interactions can lead to   topologically non-trivial excitations in a system without symmetry-breaking fields. To that end, we have first performed a detailed analysis of the symmetry properties of the non-interacting system and identified the symmetries that are responsible for the appearance of the Dirac points. The ground state is then calculated and its properties examined. In particular, we show that certain point group symmetry and $\mathcal{PT}$ sysmetry are spontaneously broken due to the presence of  particle interactions. We further show that such symmetry breaking is ultimately responsible for the creation of topologically non-trivial Bogoliubov excitations. 

\textit{Acknowledgement}.  This work is supported by the Key-Area Research and Development Program of Guangdong Province (Grant No.~2019B030330001), the National Key R\&D Program of China (Grant No.~2018YFA0307200), NSFC (Grant No.~11904417 and Grant No.~U1801661), the high-level special funds from SUSTech (Grant No.~G02206401), and Grant No.~2019ZT08X324.

\appendix
\section{Dirac point in the 2D spin-orbit coupling model}
\label{appendixA}
In this appendix, we demonstrate that the dispersion in the vicinity of the band-touching $M$ point of the non-interacting 2D spin-orbit coupling model is linear, i.e., the $M$ point is in fact a Dirac point. This can be done by using the well-known $\bm k \cdot \bm p$ perturbation method combined with the $\tilde D_4$ group symmetry analysis~\cite{dresselhaus2007group,CXLiu}. Let's write the four degenerate states at the $M$ point in the Bloch form as $\phi_{i} = e^{i\bm k_M\cdot \bm r}u_{i\bm k_M}(\bm r) $ with $i = 1,\cdots, 4$, where $\bm k_M$ is the crystal momentum at the $M$ point. Note that for convenience we have dropped the $M$ subscript in $\phi_{Mi}$ which was the notation used in the main text.  We are interested in the dispersion $\epsilon_{\bm k}$ at crystal momenta $\bm k=\bm k_M+\bm \kappa$ where $\bm \kappa$ is small. The corresponding Bloch wave function can be expanded in terms of the wave functions at $M$ point as $ u_{\bm k}(\bm r)=\sum_{i}a_i u_{i\bm k_M}(\bm r) $. The coefficients $a_i$ and the dispersion $\epsilon_{\bm k}$ can be obtained by solving the following eigen equation
\begin{equation}
\label{eq:kp}
\sum_{j}\left[\epsilon_M\delta_{ij}+\frac{\hbar}{m}\bm \kappa\cdot \bm p_{ij}\right]a_j=\epsilon_{\bm k}a_i
\end{equation}
 where $\epsilon_M$ is the energy at the $M$ point and  $\bm p_{ij}=\langle\phi_{i}|\bm p|\phi_{j}\rangle$ are the matrix elements of the single particle momentum operator $\bm p$.
 
We now make use of the $\tilde D_4$ group symmetry to determine the selection rules for the matrix elements $\bm p_{ij}$. As explained earlier in the main text, the wave functions at the $M$ point can be divided into two sets $\{\phi_1,\phi_2\}$ and $\{\phi_3,\phi_4\}$, each of which can be used to construct a $2D$ irreducible representation of the $\tilde D_4$ group, denoted by $E_2$ and $E_3$ respectively. The selection rules of $\bm p_{ij}$ can be obtained from the knowledge of how these states transform under the group symmetry operations. 

 In fact, the transformations of states $\{\phi_1,\phi_2\}$ under the group symmetry operations are the same as the linear basis of $E_2$, namely $\left\{ \left(\begin{smallmatrix} x-iy\\0\end{smallmatrix}\right), \left(\begin{smallmatrix} 0\\x+iy\end{smallmatrix}\right)\right\}$. 
Let's take the operation $\tilde{r}_2=e^{-i\frac{\pi}{2}\sigma_z}r_2$ as an example. Since $r_2\{x,y\}=\{-x,-y\}$,  we have
 \begin{equation}
 	\tilde{r}_2\biggl\{ \biggl(\begin{matrix} x-iy\\0\end{matrix}\biggr),\biggl(\begin{matrix}0\\x+iy\end{matrix}\biggr)\biggr\}=\biggl\{ i\biggl(\begin{matrix} x-iy\\0\end{matrix}\biggr),-i\biggl(\begin{matrix}0\\x+iy\end{matrix}\biggr)\biggr\},
 \end{equation}
which leads to 
\begin{equation}
\tilde{r}_2\{\phi_{1},\phi_{2}\}=\{i\phi_{1},-i\phi_{2}\}.
\end{equation} 
  The transformations of states $\{\phi_3,\phi_4\}$ under the group symmetry operations are the same as the linear basis of $E_3$, i.e., $\left\{ \left(\begin{smallmatrix} x+iy\\0\end{smallmatrix}\right), \left(\begin{smallmatrix} 0\\x-iy\end{smallmatrix}\right)\right\}$, then we have
$\tilde{r}_2\{\phi_{3},\phi_{4}\}=\{i\phi_{3},-i\phi_{4}\}$. From these transformation properties it is straightforward to show that the matrix element $\bm p_{11}$ satisfies
\begin{align}
	\langle \phi_1|\bm p|\phi_1\rangle &=\langle \phi_{1} | \tilde{r}_{2}^{\dagger} \tilde{r}_{2}\bm p \tilde{r}_{2}^{\dagger}\tilde{r}_2 | \phi_1\rangle =\langle i\phi_1|(-\bm p)|i\phi_1\rangle\notag \\ 
	{} & = -\langle \phi_1|\bm p|\phi_1\rangle = 0, 
\end{align}
where we used the fact that $\tilde{r}_2\bm p\tilde{r}_2^{\dagger}=-\bm p$. 
By the same token one can check that $\bm p_{13},\bm p_{24}$ and all the diagonal elements $\bm p_{nn}$ vanish as well.  

Next we show that $\bm p_{12}  = \bm p_{34} = 0 $ from the $\mathcal{PT}$ symmetry. Denoting $\mathcal C\equiv \mathcal{PT}=i\sigma_{y}\mathcal{KI}$, where $\mathcal I$ is the space inversion operator and $\mathcal K$ is the complex conjugation operator, we have 
\begin{equation}
	\mathcal{C}\{\phi_1,\phi_2\}=\{\phi_2,-\phi_1\},\quad \mathcal{C}\{\phi_3,\phi_4\}=\{\phi_4,-\phi_3\}.
\end{equation}
Since $\mathcal C$ ia an anti-unitary operator, we find
\begin{align}
	\langle \phi_1|\bm p|\phi_2\rangle &=\langle \mathcal C\phi_{1} | \mathcal C(\bm p \phi_2)\rangle^{*}=\langle \phi_{2} | \bm p (-\phi_1)\rangle^{*} \notag \\ 
	{} & = -\langle \phi_1|\bm p|\phi_2\rangle=0.
\end{align}
Similarly, we can also show $\bm p_{34}=0$. 

The remaining nonzero matrix elements are  $\bm p_{14}, \bm p_{23}$ and their hermitian conjugates. By considering the operations $\tilde{s}_2=e^{-i\frac{\pi}{2}\sigma_y}s_2$ and $\tilde{r}_1=e^{-i\frac{\pi}{4}\sigma_z}r_1$,
we find that  the nonzero elements satisfy 
\begin{equation}
\label{eq:MatrixElement}
	\langle \phi_1| p_x|\phi_4\rangle=\langle \phi_2| p_x|\phi_3\rangle=i\langle \phi_1|p_y|\phi_4\rangle=-i\langle \phi_2| p_y|\phi_3\rangle.
\end{equation}
Denoting $t=\frac{\hbar}{m}\langle \phi_1| p_x|\phi_4\rangle$, which can always be made real by choosing a proper gauge. We can write the Hamiltonian near $M$ point as
\begin{equation}
\label{eq:kph}
	h(\bm k_M+\bm \kappa)=\epsilon_M\mathbb{I}+t\kappa_x \sigma_x\otimes\sigma_x+t\kappa_y\sigma_x\otimes\sigma_y.
\end{equation}
Diagonalizing Eq.~(\ref{eq:kph}) immediately leads to the linear dispersion
\begin{equation}
	\epsilon_{\bm k_M+\bm \kappa}-\epsilon_M=\{-t|\bm \kappa|,-t|\bm \kappa|,t|\bm \kappa|,t|\bm \kappa|\}.
\end{equation}

\bibliography{BibTex3.bib}
\end{document}